\documentclass[a4paper,11pt] {article}
\usepackage{epsfig, graphicx}
\usepackage{float}
\usepackage{amsmath}
\usepackage{latexsym}
\usepackage{epsfig, graphicx} 
\usepackage{graphics}

\begin{document}
\title{Toward a quantitative analysis of virus and plasmid trafficking in cells }
\author{Thibault Lagache (+), Emmanuel Dauty(+) and David Holcman (+,*) \\
 \textit{*: Department of Mathematics, Weizmann Institute of
Science} \\\textit{ Rehovot 76100, Israel}  \\ \textit{  +:  D\'epartement de
Math\'ematiques et de Biologie, Ecole Normale Sup\'erieure }\\ \textit{46 rue
d'Ulm 75005 Paris, France}\\ D. H. is supported by
the program ``Chaire d'Excellence'' \\ This research is supported by
the grant Human Frontier Science \\Program 0007/2006-C.}

\begin{abstract}
Intracellular transport of DNA carriers is a fundamental step of
gene delivery. We present here a theoretical approach to study
generically a single virus or DNA particle trafficking in a cell
cytoplasm. Cellular trafficking has been studied experimentally
mostly at the macroscopic level, but very little has been done so
far at the microscopic level. We present here a physical model to
account for certain aspects of cellular organization, starting
with the observation that a viral particle trajectory consists of
epochs of pure diffusion and epochs of active transport along
microtubules. We define a general degradation rate to describe the
limitations of the delivery of plasmid or viral particles to the
nucleus imposed by various types of direct and indirect hydrolysis
activity inside the cytoplasm. Following a homogenization
procedure, which consists of replacing the switching dynamics by a
single steady state stochastic description, not only can we study
the spatio-temporal dynamics of moving objects in the cytosol, but
also estimate the probability and the mean time to go from the
cell membrane to a nuclear pore. Computational simulations confirm
that our model can be used to analyze and interpret viral
trajectories and estimate quantitatively the success of nuclear
delivery.
\end{abstract}

\maketitle

\section*{Introduction}
The study of the motion of many particles inside a biological cell
is a problem with many degrees of freedom and a large parameter
space. The latter may include the different diffusion constants of
the different species, velocities along microtubules, their
number, the geometry of cell and nucleus, the number and sizes of
nuclear pores, the various degradation factors, and so on. The
experimental and numerical exploration of this multi-dimensional
parameter space is limited perforce to a small part thereof, due
to the great complexity of the biological cell. A great reduction
in complexity is often achieved by coarse-graining the complex
motion by means of effective equations and their explicit
analytical solutions, which is the approach we adopt in this
letter. We are specifically concerned with finding a concise
description of virus and plasmid trafficking in the cell
cytoplasm.

Early vesicle trafficking studies revealed the complex secretion
pathways \cite{Palade}, whereas much more recent studies of
natural (viruses) \cite{Greber,charneau,Seisenberger} and
synthetic (amphiphiles) DNA carriers \cite{Zuber} uncover details
of the cellular pathways and the complexity of cellular infection.
Viruses invade mammalian cells through multistep processes, which
begin with the uptake of particles, cytoplasmic trafficking, and
nuclear import of the DNA. However, cytoplasmic trafficking
remains a major obstacle to gene delivery, because the cytosolic
motion of large DNA molecules is limited by physical and chemical
barriers of the crowded cytoplasm \cite{Verkman,Dauty}. Whereas
molecules smaller than 500kDa can diffuse, larger cargos such as
viruses or non-viral DNA particles, require an active transport
system \cite{sodeik}. Viral infection is much more efficient than
gene transfer using polymers- or lipids-based vectors, where a
large amount of endocytozed DNA (typically over 100.000 copies of
the gene) is required to produce a cellular response, while only a
few copies seem to be necessary in the case of viruses.

Two recent studies \cite{nedelec,pangarkar} showed that
microtubules shape the distribution of molecular motors and
vesicle trafficking inside the cell cytoplasm by means of a
combination of experiments and numerical simulations. The mean
concentration of viral species was analyzed in \cite{Dinh} by
means of the mass-action law. The mechanism of DNA transport in
the cytoplasm, however, is still an open question. We propose here
a coarse-grained reduced description of viral trafficking and
compare it to plasmid diffusion. Specifically, we are interested
in the probability $p_N$ and the mean time $\tau_N$ for a DNA
carrier or a virus to get from the cell membrane to a small
nuclear pore. The evaluation of these quantities calls for a
quantitative approach to the description of particle trajectories
at an individual level and also, to quantify the role of the cell
organization and the signaling processes involved in viral
infection.

\begin{figure}
\center
\includegraphics[width=5cm]{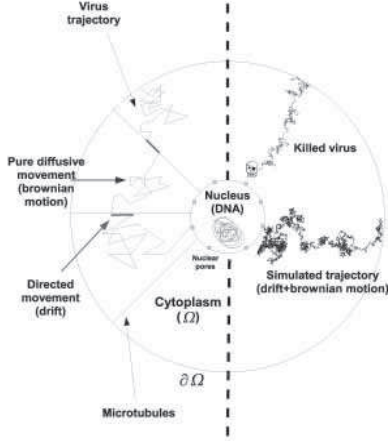}
\caption{On the left-side of the idealized cell, a
real trajectory consists of intermittent Brownian and drift
epochs, whereas on the right-side, we show two simulated
homogenized trajectories. In one of them, the virus arrives alive
to a nuclear pore, while in the other, it is killed inside the
cytoplasm. The round dots on the nucleus surface represent the
nuclear pores. }\label{FIGURE1}
\end{figure}

We start with the observations that a viral movement can be
described as a combination of intermittent switches between pure
Brownian diffusion and active transport along microtubules (figure
\ref{FIGURE1}), while DNA motion can be characterized as pure Brownian.
We also account for multiple factors involved in degradations,
such as hydrolyzation, destruction in lysosomes, or any other
factors that prevent irreversibly the particle from reaching a
nuclear pore. This degradation process is modeled as killing with
a time-independent rate $k$. We use the overdamped Langevin
dynamics with killing to describe the viral or DNA motion and use
Fokker-Planck-type equations to obtain asymptotic approximations
of $p_N$ and $\tau_N$ in the limit of large and small $k$. We
compute the mean time the first among many independent viruses
reaches a small nuclear pore. Brownian simulations confirm the
validity of the analytical analysis. The present approach is a
first attempt to develop a theoretical tool for the analysis of
virus dynamics and, hopefully, for the study of trafficking of
synthetic vectors, a necessary step toward gene delivery.
\par

{\noindent \bf Modeling Viral or DNA trajectories}
We model viral trajectories as a collection of pieces, each of
which is characterized either as directed movement along
microtubules or pure Brownian motion
\cite{Greber,charneau,Seisenberger}. In contrast, DNA motion in
the cytoplasm can be adequately described as pure Brownian motion
\cite{Dauty}. Particles moving inside the cell are reflected at impermeable
surfaces and are absorbed at nuclear pores.
A virus travels on microtubules as long as it binds to {\bf a}
motor. The three- or two-dimensional position of a particle, $X(t)$,
is described by the coarse-grained stochastic dynamics
 \begin{equation}
 \dot{X}= \left\{\begin{array}{l} \sqrt
{2D }
\dot {w} \quad \mbox{for a free particle }\\ \\
{\bf V}(t) \quad \mbox{for a bound particle }
\end{array}\right.,\label{eq1}
\end{equation}
where ${w}$ is a $\delta$-correlated standard white noise and
${\bf V}(t)$ is a time-dependent velocity along a microtubule. The
velocity ${\bf V}(t)$ can be either  positive or negative,
depending on whether a viral particle binds to a dynein or to a
kinesin motor. However, it is not clear what regulatory mechanisms
is involved in such a choice
\cite{Welte}.

{\noindent \bf Mathematical description of a viral trajectory in
the cytoplasm.}
We consider the trafficking of a viral particle from an endosome
or the cell membrane to a small nuclear pore. The cell cytosol is
a bounded spatial domain $\Omega$, whose boundary $\partial
\Omega$ is the external membrane $\partial \Omega_{ext}$ and the nuclear envelope.
Most of the nuclear membrane consists of a reflecting boundary
$\partial N_r$, except for small nuclear pores $\partial N_a$,
where a viral particle can enter the nucleus. We assume that a
viral particle that reaches a pore is instantly absorbed, so that
this boundary is purely absorbing for trajectories. The ratio of
the surface areas is assumed small,
 \begin{equation}
\varepsilon=\frac{|\partial N_a|}{|\partial \Omega|} \ll 1.\label{smallnp}
 \end{equation}

{\noindent \bf Homogenization of viral trajectory.}
To replace the intermittent dynamics between free diffusion and the
drift motion along microtubules{\bf,} described in equation
(\ref{eq1}), we use the precise calibration procedure described in
\cite{thibault}. In this homogenization procedure, the motion is
described by the overdamped limit of the Langevin equation
 \begin{equation}  d\mathbf{X} =
\mathbf{b}\left(\mathbf{X}\right)\,dt + \sqrt{2D}\,d\mathbf{W},\label{eq2}
 \end{equation}
where $D$ is the diffusion constant and $\mathbf{b}(\mathbf{X})$
represents the steady state drift. Because all microtubules
starting from the cell surface converge to the centrosome, a
specialized organelle located nearby the cell nucleus (figure
\ref{FIGURE1}), we choose in the first approximation a
radially symmetric effective drift $\mathbf{b}(\mathbf{X})$ converging to
the nucleus. This approximation can be justified by the study
\cite{charneau}, where viral trajectories move around the nucleus
surface. Although viruses move bidirectionally on microtubules,
the overall movement is directed toward the nucleus, thus we only
consider here this average motion \cite{Welte}. The homogenized
drift in (\ref{eq2}) becomes
\begin{equation} \label{eq2b}
\mathbf{b}=-B\frac{\mathbf{X}}{|\mathbf{X}|}
\end{equation}
where $B$ is a constant amplitude, which depends on many
parameters, such as the density of microtubules, the binding and
unbinding rates and the averaged velocity of the directed motion
along microtubules \cite{thibault}.

{\noindent \bf From trajectory description to the probability and
mean arrival time.}
{\bf V}iral killing, immobilization or rejection out of the cell
and naked DNA degradation by nucleases, are coarse-grained into a
steady state killing rate $k$. To derive asymptotic expressions
for the probability $P_N$, that a DNA carrier (single virus or
DNA) arrives to a small nuclear pore alive and for the mean time
$\tau_N$, we use the approximation (\ref{smallnp}). The asymptotic
estimates depend on the diffusion constant $D$, the amplitude of
the drift $B$, and $k$. These computations are based on the small
hole theory \cite{HS}, which describes a Brownian particle
confined to a bounded domain by a reflecting boundary, except for
a small absorbing window, through which it escapes. The domain
$\Omega$ contains a spherical nucleus of small radius $\delta$.
The survival probability density function (SPDF) $p(\mathbf{x},t)$ to find
the virus or naked DNA alive inside the volume element $\mathbf{x}+d\mathbf{x}$ at
time $t$ is given by
 \begin{equation}
p(\mathbf{x},t) d\mathbf{x}= Pr\{X(t) \in \mathbf{x}+ d \mathbf{x}, \tau^k>t ,\tau^a>t| p_i \},
 \end{equation}
where $\tau^a$ is the first passage time of a live DNA carrier to
the absorbing boundary $\partial N_a $, $\tau^k$ is the first time it is
hydrolyzed, and $p_i$ is the initial distribution. The SPDF
$p(\mathbf{x},t)$ of the motion (\ref{eq2}) is the solution of the mixed
initial boundary value problem for the Fokker-Planck equation
(FPE)
\cite{Schuss}
 \begin{eqnarray*}
\frac{\partial p}{\partial t}(\mathbf{x},t)&=&D\Delta p(\mathbf{x},t) -
\nabla\cdot\mathbf{b}(\mathbf{x})p(\mathbf{x},t)
 -k p(\mathbf{x},t) \nonumber\\
&&\nonumber\\
 p(\mathbf{x},0) & = & p_i(\mathbf{x})\quad\mbox{for}\quad\mathbf{x}\in \Omega\label{FPE}\\
&&\nonumber\\
p(\mathbf{x},t)&=&0\quad\mbox{for}\quad\mathbf{x}\in\partial N_a\nonumber\\
&&\nonumber\\
\mathbf{J}(\mathbf{x},t)\cdot\mathbf{n}_{\mathbf{x}}&=& 0 \quad \mathbf{x} \in \partial N_r \cup \partial
\Omega_{ext} \label{boundary-c2},
 \end{eqnarray*}
where $\mathbf{n}_{\mathbf{x}}$ is the unit outer normal at a boundary point $\mathbf{x}$.
The flux density vector $\mathbf{J}(\mathbf{x},t)$ is defined as
 \begin{equation}
\mathbf{J}(\mathbf{x},t)=-D \nabla p(\mathbf{x},t) +\mathbf{b}(\mathbf{x})p(\mathbf{x},t).\label{Ji}
 \end{equation}
The probability $P_N$ that a live DNA carrier arrives at the
nucleus is $P_N = Pr\{\tau^a<\tau^k\}$ \cite{HMS}. This
probability can be expressed in terms of the SPDF \cite{HMS} by
 \begin{eqnarray*}
P_N=1-Pr\{\tau^a>\tau^k \} &=&
1-\int_{\Omega}k(\mathbf{x})\tilde{p}(\mathbf{x})\,d\mathbf{x},
 \end{eqnarray*}
where $ \tilde{p}(\mathbf{x})= \int_0^\infty p(\mathbf{x},t)\,dt$ is the solution
of equation
 \begin{equation}q 
D\Delta \tilde{p}(\mathbf{x})-\nabla\cdot\mathbf{b}(\mathbf{x})\tilde p(\mathbf{x})
-k(\mathbf{x})\tilde{p}(\mathbf{x})= -p_i(\mathbf{x})\quad\hbox{for}\quad\mathbf{x}\in\Omega
 \end{equation}q
with the boundary conditions (\ref{boundary-c2}). Using the pdf of
the time to absorption, conditioned on the event that the DNA
carrier escapes alive $\Pr\{\tau^a<t\,|\,\tau^a<\tau^k\}$, we
define the conditional mean time to absorption as
 \begin{eqnarray*}
\tau_{N}&=&E[\tau^a \,|\,\tau^a<\tau^k] =\int_0^\infty
{(1-\Pr\{\tau^a <t\,|\,\tau^a<\tau^k)}\,dt.
 \end{eqnarray*}
Following the computations of \cite{david}, we get
 \begin{equation}
\tau_N = \frac{\int_{\Omega} \tilde{p}(\mathbf{x}) d\mathbf{x} -\int_{\Omega}
k(\mathbf{x}) q(\mathbf{x})d\mathbf{x}}{1-\int_{\Omega}k(\mathbf{x})\tilde{p}(\mathbf{x})\,d\mathbf{x}},\label{ey},
 \end{equation}
where
 \begin{equation}
 q(\mathbf{x} )=\int_{0}^{\infty} s\tilde{p}(\mathbf{x},s)\,ds
 \end{equation}
satisfies \cite{david}
 \begin{equation} \label{FPEpi}
-\tilde{p} =D\Delta q(\mathbf{x})-\nabla\cdot\mathbf{b}q] -k q\quad
\hbox{for}\quad\mathbf{x}\in \Omega
 \end{equation}
with boundary conditions (\ref{boundary-c2}).

\begin{figure}
\includegraphics[width=4.3cm]{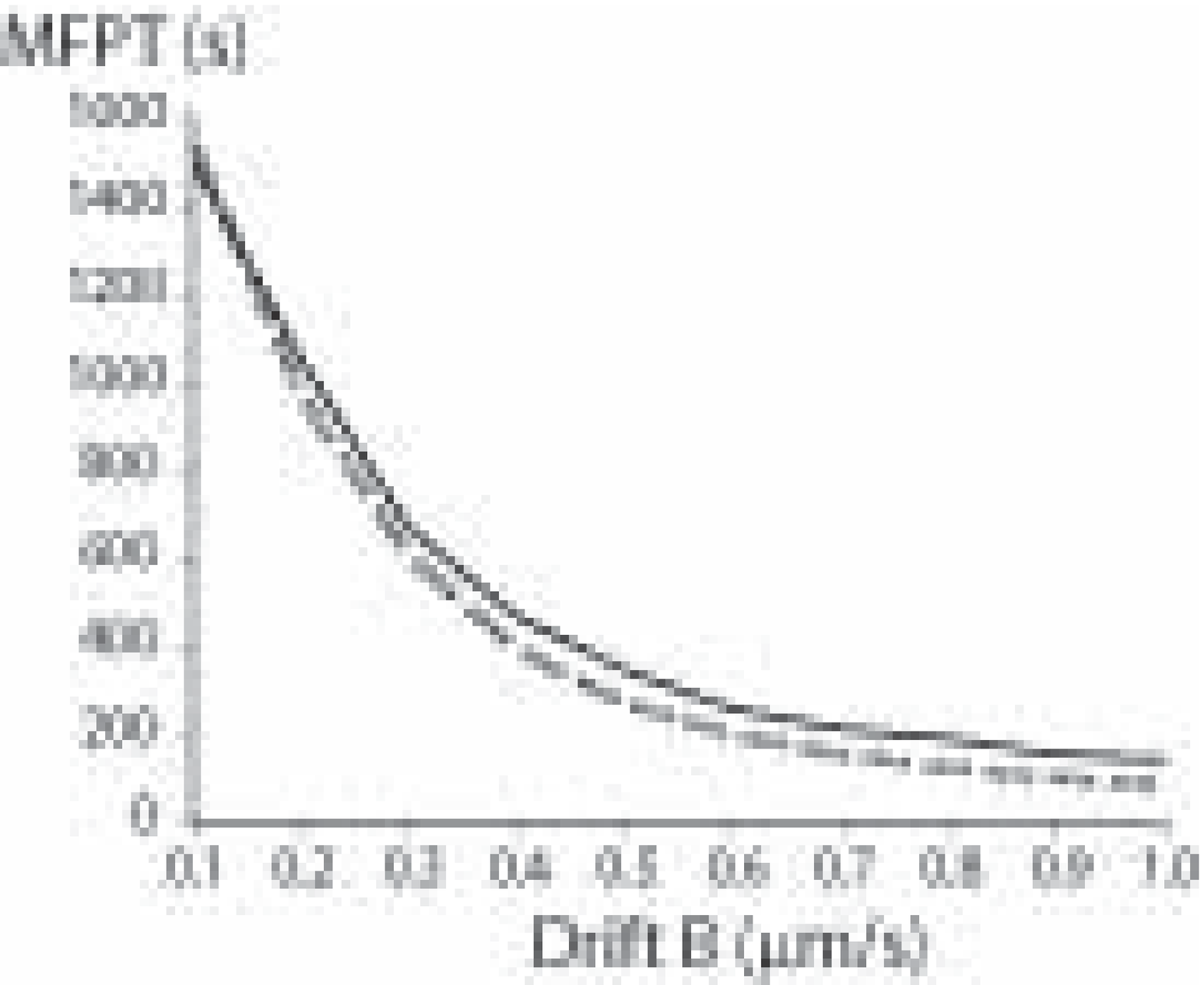}\hfill
\includegraphics[width=4.3cm]{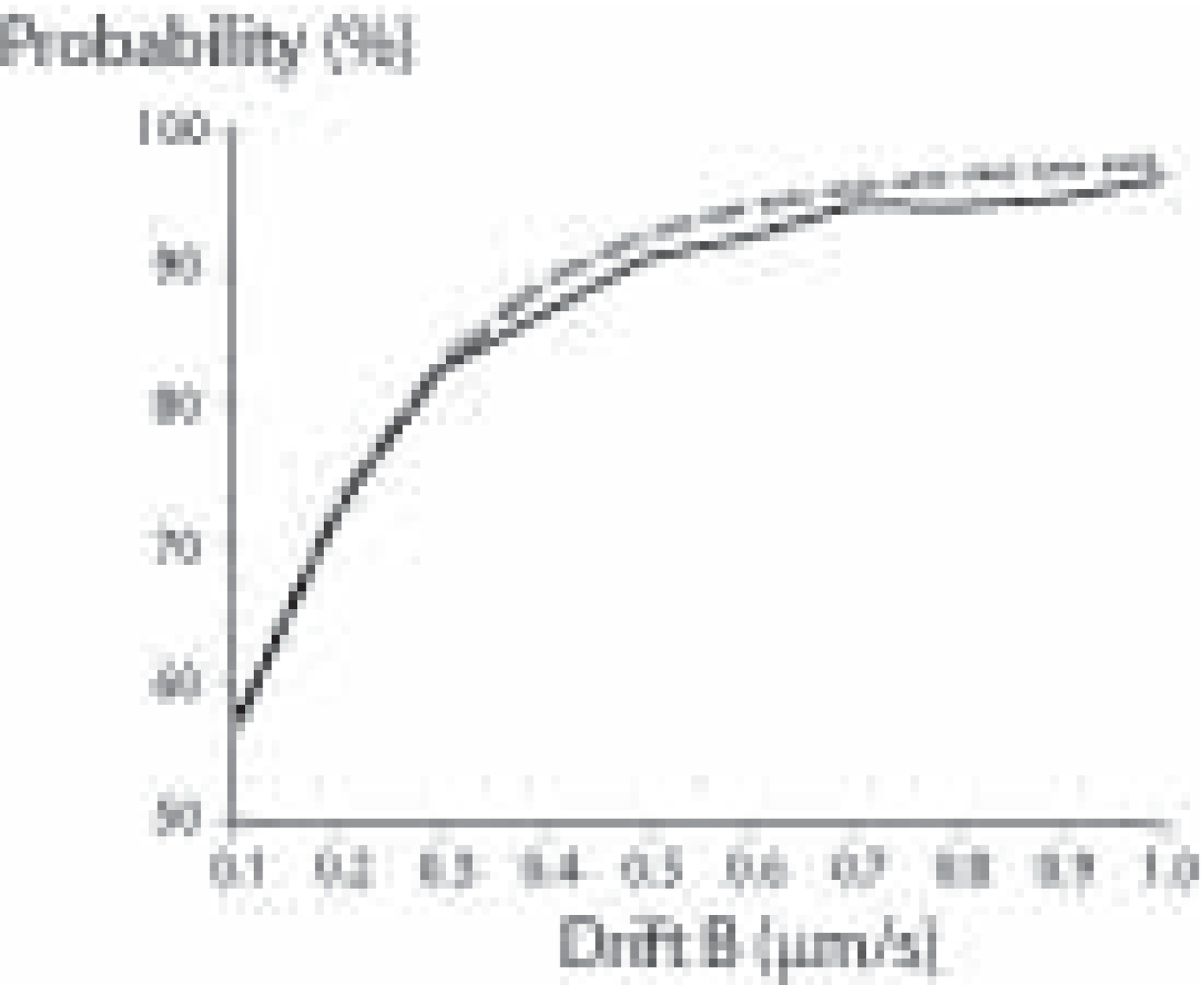}\hfill
\includegraphics[width=4.3cm]{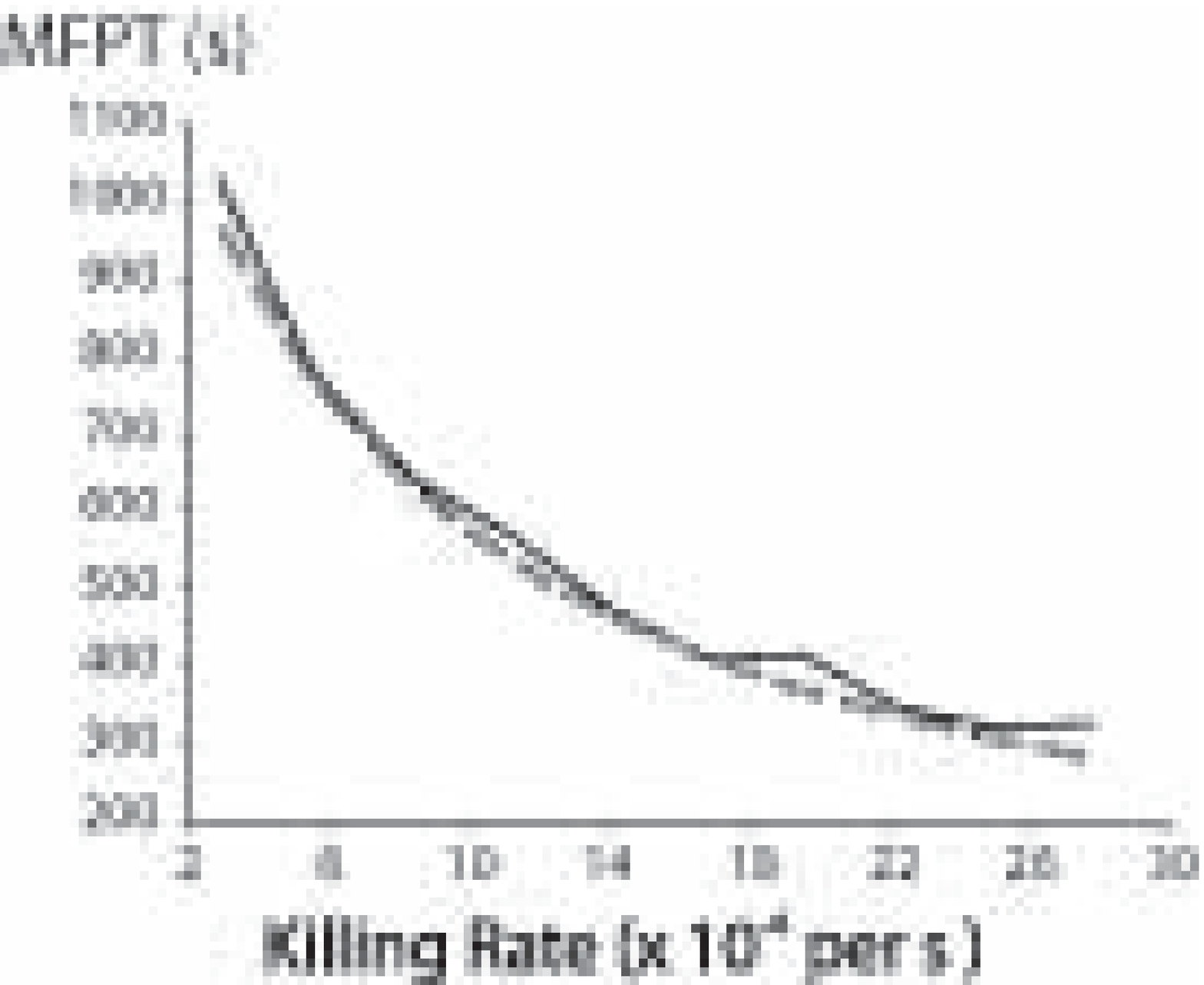}\hfill
\includegraphics[width=4.3cm]{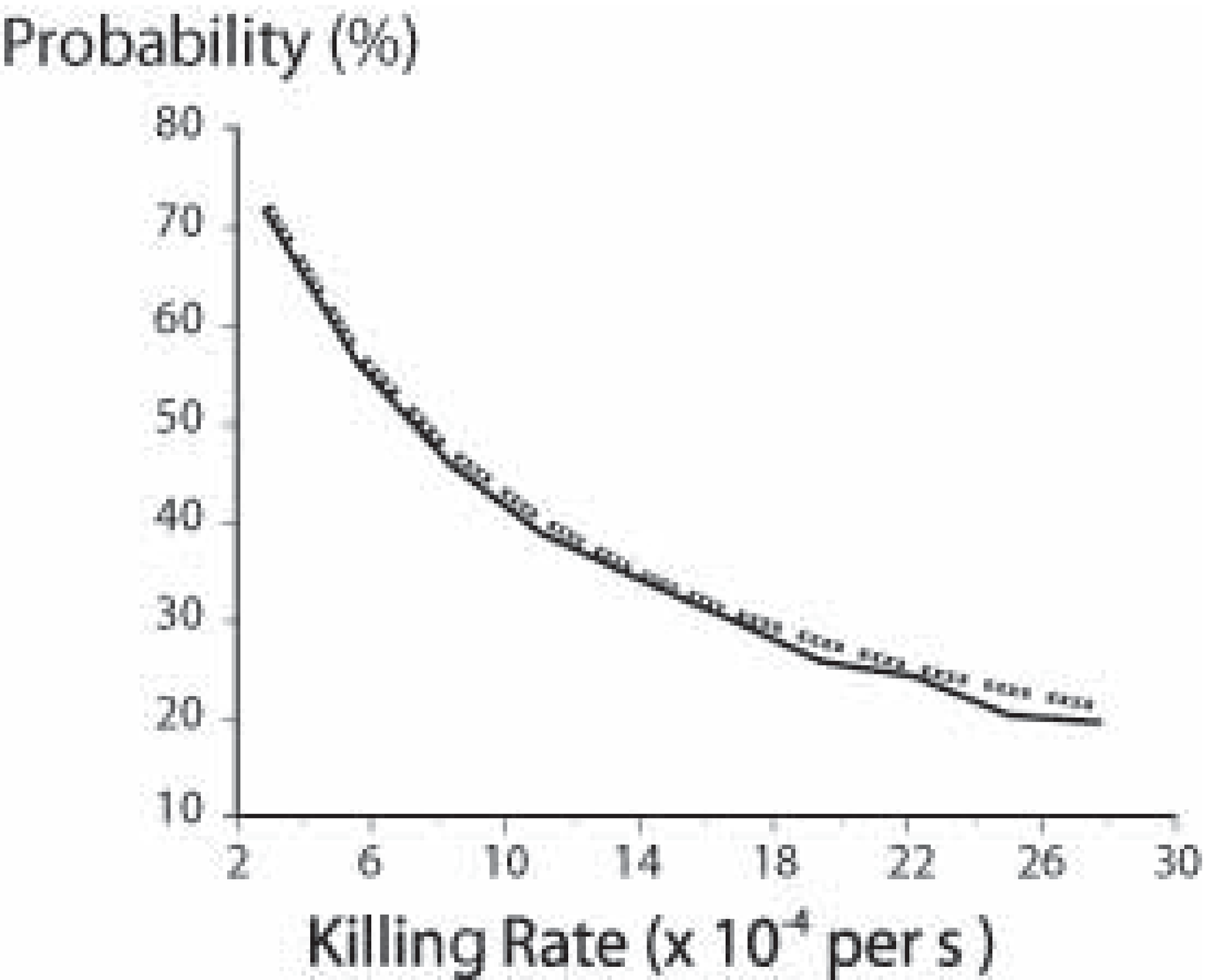}
\caption{\small{ MFPT (top left) and the arrival probability (top
right) for increasing values of the drift with $k=\frac{1}{3600}
s^{-1}$ and for increasing values of the steady state killing rate
($B=0.2\mu m s^{-1}$) (bottom). $2000$ trajectories are simulated,
theoretical values and simulated ones are drawn with dashed and
solid lines, {respectively}. $R=20\mu m$; $\delta=4\mu m$;
$\eta=\frac{\pi}{12}\delta=1.05 \mu m$; $D=1.3\mu m^2
s^{-1}$\cite{Seisenberger}; $n=1$.}}\label{FIGURE2}
\end{figure}

{\noindent \bf Asymptotic expressions: the plasmid case.}
To obtain explicit expression for $P_N$ and $\tau_N$ for a nucleus
containing $n$ well separated small holes (nuclear pores) on its
surface, we consider first a killing rate $k$ smaller than the
diffusion constant $D$. The asymptotic analysis for naked DNA
$(\mathbf{b}=\mathbf{0})$ leads to \cite{david}
 \begin{equation}
P_N= \frac{1}{1+\frac{|\Omega |\tilde k}{4nD\eta }} , \quad \tau_N = \frac{\left(\frac{|\Omega|}{4D\eta n
}\right)}{1+\left(\frac{|\Omega |\tilde
k}{4nD\eta}\right)},\label{tau-ff}
 \end{equation}
where $\tilde k=\frac{1}{|\Omega |}\int_\Omega k(\mathbf{x})\,d\mathbf{x}$,
and $\eta$ is the radius of a small absorbing disk. Formula
(\ref{tau-ff}) does not depend on the specific shape of the
killing rate $k$, but rather on its integral. We compare this
asymptotic formula with Brownian simulations obtained for
parameters $R=20\mu m$; $\delta=\frac{R}{5}$;
$\eta=\delta\frac{\pi}{12}=1.05\mu m$; $k=\frac{1}{3600}s^{-1}$
\cite{Lechardeur}; $D=0.02 \mu m^2 s^{-1}$ \cite{Dauty}; $n=1,$
({a single} big hole). This simulation corresponds to a cell
{with} $2\%$ of the nuclear surface {occupied} by a large nuclear
{pore, or equivalently,} to a simulation with $n=2000$ pores of
radius $25 nm$\cite{maul}. Because formula (\ref{tau-ff}) depends
only on the product $n \eta$, both simulations give the same
result.
\begin{center}
\begin{tabular}{|c|c|c|}
\hline
Time and Probability & $\tau_N$ & $ P_N$ \\
\hline
Theoretical values & \hbox{ } $3567 s$ \hbox{ }& \hbox{ }$0.90\%$\hbox{ } \\
\hline
Simulated values (2000 particles.) & \hbox{ } $3564s$ \hbox{ } &\hbox{ } $0.97\%$ \hbox{ } \\
\hline
\end{tabular}
\end{center}
When $k(\mathbf{x})$ is much larger compared to diffusion, a boundary
layer analysis leads to the asymptotic expression
\begin{eqnarray*}
P_N = \frac{n\eta |\partial \Omega|}{|\Omega|}\sqrt{\frac{D}{k_0}},
\end{eqnarray*}
where we assume that the smooth killing {rate is a constant $k_0$
in the neighborhood of the nuclear surface.}
{\noindent \bf Asymptotic expressions: the virus case.}
For a virus trajectory governed by equation (\ref{eq2}), with a
constant {scalar drift $B$}, the leading order term of the
probability and the mean time are given by
\cite{david}
 \begin{equation} \label{tau-fff}
 P_N =
\frac{1}{\frac{\pi}{nD\eta}\left(\frac{D}{B}\delta^2+2\left(\frac{D}{B}\right)^2\delta
+2\left(\frac{D}{B}\right)^3\right)k+1} \\
\tau_N =
\frac{\frac{\pi}{nD\eta}\left(\frac{D}{B}\delta^2+2\left(\frac{D}{B}\right)^2\delta+
2\left(\frac{D}{B}\right)^3\right)}{\frac{\pi}{nD\eta}\left(\frac{D}{B}\delta^2+2\left(\frac{D}{B}\right)^2
\delta+2\left(\frac{D}{B}\right)^3\right)k+1}. \nonumber
 \end{equation}
These asymptotic formulas show that the main contribution to the
probability and the mean time comes from a boundary layer located
near the nucleus surface. { The killing rate $k$ in this case} is
the averaged value of the killing field in that boundary layer
\cite{david}. In figure \ref{FIGURE2}, we compare the probability to
arrive {alive at the pore} and the mean {arrival} time for several
values of the drift and the constant killing rate.
\begin{figure}
\centering {\begin{minipage}[c]{5.5cm}
\includegraphics[width=5.5cm]{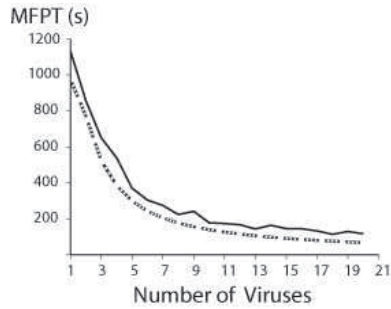}
\end{minipage}
\begin{minipage}[c]{3cm}
\caption{\small{MFPT of the first virus. $300$ trajectories are
simulated (solid line). $R=20\mu m$; $\delta=4\mu m$;
$\eta=\frac{\pi}{12}\delta=1.05 \mu m$; $D=1.3\mu m^2 s^{-1}$; $n=1$
and $B=0.2 \mu m s^{-1}$.}} \label{FIGURE3}
\end{minipage}}
\end{figure}
For {a large number} of microtubules, the drift $B$ equals the
apparent velocity \cite{thibault} ($10\%$ \cite{Suomalainen} of
the minus end velocity, approximatively equal to $2 \mu m/s$
\cite{Seisenberger}). Using formula (\ref{tau-fff}), we can now
predict the effect of changing the effective drift $B=0.2$ by $\pm
30\%$: increasing the drift leads to a probability $P^{+30\%}_N
=0.80$ and a mean time $\tau^{+30\%}_N=731s$, while reducing the
drift gives $P^{-30\%}_N =0.64$ and $\tau^{-30\%}_N=1293s$. We
conclude that decreasing the drift increases the time by $33\%$
($\tau_N=974s$) and decreases the probability by $12\%$
($P_N=0.73$), while increasing the drift, reduces the time by $22\%$
and increases the probability by $10\%$. These results show the
nonlinear effect of the drift. In a biological context, decreasing
the drift can be implemented by disrupting the microtubule network.

{\noindent \bf Mean first passage time of the first virus to reach
the nucleus.}
When $M$ viruses enter a cell, the {number} $M_a$ {of live viruses
arriving at} the nucleus is given by $M_a = P_{N}M$. The {mean
time the first live virus arrives at a nuclear pore is given by}
 \begin{equation}
\tau_{first}(M) = \frac{\tau_N}{1-\mathbf{x}i^M}\left( \sum_{k=0}^{M-1}
\frac{\mathbf{x}i^k}{M-k} +
\mathbf{x}i^M\sum_{k=1}^{M}\frac{\left(-1\right)^k}{k}\right)
 \end{equation}
where $\mathbf{x}i = 1-P_N$. Finally, {\bf asymptotic} expansions give
 \begin{eqnarray*}
\tau_{first} &\approx& \left\{\begin{array}{l}
\frac{\tau_N}{M}\left(1+\frac{M}{M-1}\mathbf{x}i\right) \hbox{ for } \mathbf{x}i \ll 1 \\ \\
\frac{ln(\frac{M}{2})}{M}\frac{\tau_N}{1-\mathbf{x}i} \hbox{ for } \mathbf{x}i
\approx 1 \hbox{ and }M >>1.
\end{array}\right.
 \end{eqnarray*}
{The theoretical results are compared with Brownian simulations in
figure} \ref{FIGURE3}.

{The closed form expressions \ref{tau-ff}-\ref{tau-fff} facilitate
the exploration of the multi-dimensional parameter space of
cellular delivery of both DNA and virus trafficking}. Cytoplasmic
trafficking is a limiting step of gene delivery. {Elucidating
viral motion in the cytoplasm may provide a quantitative tool for
the improvement and optimization of delivery of synthetic
vectors}. {The present approach can provide a resource for
optimizing the design of synthetic vectors and for the analysis of
the parameters} of viral infection.

\end{document}